\documentclass{article}
\usepackage{graphicx}
\usepackage{amssymb,amsthm,amsmath}
\usepackage{xcolor,hyperref,fancyhdr}
\usepackage{stackengine} % subfigure labeling

\makeatletter

\title{Variations on the Three-Sphere: Laves' Labyrinth Lopped}

\renewcommand\@date{{
    \vspace{-\baselineskip}
    \large\centering
    \begin{tabular}{@{}c@{}}
        Lauren Niu\textsuperscript{1,2} \\
        \normalsize lniu@sas.upenn.edu
    \end{tabular}
    \qquad \qquad
    \begin{tabular}{@{}c@{}}
        Randall D. Kamien\textsuperscript{1,3} \\
        \normalsize kamien@upenn.edu
    \end{tabular}
    
    \bigskip
    
    {\small \it
        \textsuperscript{1}Department of Physics and Astronomy, University of Pennsylvania, Philadelphia, PA\par
        \textsuperscript{2}arXiv, Cornell University, New York, NY\par
        \textsuperscript{3}Department of Mathematics, University of Pennsylvania, Philadelphia, PA
    }
    
    \bigskip
    
    {\small \today}
}}
\makeatother

\begin{document}

\maketitle

\begin{abstract}
Inspired by the structure of $srs$ Laves networks in $\mathbb{R}^3$ that underpin the celebrated gyroid surface, we construct a Laves network of identical three-coordinated vertices on $S^3$ with double-twist. This network is a subset of the vertices and edges of the 600-cell, and can be viewed as a bipartite graph of disjoint 24-cell vertices inscribed in the 600-cell. We describe mutually entangled realizations of this network on $S^3$, and describe their relation to the well-known $srs$ Laves network structure in $\mathbb{R}^3$.
\end{abstract}

\bigskip

\section*{Introduction}

It is axiomatic that, without Euclid, we would not be studying non-Euclidean geometry. Though Euclid viewed our world as $\mathbb{R}^3$ \cite{euclid-300elements}, the study of materials typically occurs on the three-torus $T^3$; although the space is flat, its periodicity enables the wonders of the Fourier transform \cite{bragg1915x, brillouin1953wave, o2020crystal}. Different compact 3-manifolds are of interest for any number of reasons, but here we fixate on the question: what is special about the flatness of space in comparison to its dimension? While the usual mantra of differential geometry is ``curved space can be approximated locally by flat space'' the unusual converse holds as well: ``locally, a piece of flat space can be approximated by a bit of curved space.'' This flipped narrative has led to a deeper understanding of sphere packing \cite{coxeter1958close, frank1958complex, frank1959complex, nelson1989polytetrahedral, sadoc1999geometrical, virial, modes2008geometrical}, the liquid-crystalline blue phase \cite{sethna1983relieving,coates1973optical, wright1989crystalline}, and, recently, the packing of polyhedra \cite{schonhofer2023rationalizing}. Whether profound or pedigreed, this notion has become part of the standard lore, no stranger than the consideration of the Ising model in 3.99 dimensions \cite{wilson1972critical}.

Here we note that another structure that features double-twist is the celebrated Laves network in $\mathbb{R}^3$ or $T^3$ (also known as the $srs$ network, $K_4$ crystal, or triamond structure), a complex and chiral yet highly symmetric structure that can pack space with identical or enantiomorphic copies of itself \cite{laves1932klassifikation, coxeter1955laves, hyde2008short}. To see this, recall that the $\mathbb{R}^3$ Laves network is both vertex- and edge-transitive; each vertex has three coplanar neighbors, connected to each by edges that are exactly $120^\circ$ apart. Along each edge of the network, the planes spanned by the edges emerging from the two connected vertices twist by $\pm \arccos\left(1/3\right) \approx 70.5^\circ$ around the edge direction, and the sign of this twist is consistent for every edge in the network. These planes correspond to a line field at each vertex and, because of the regularity of the twist along each edge, we see that the line field twists in both directions around each vertex: double twist!

The separating surface between a Laves network and its mirror image is the gyroid \cite{schoen1970infinite}, an achiral triply periodic minimal surface that acts as a scaffold for a common ground state configuration of diblock copolymer systems and is observed in a variety of mesoscopic self-assembled materials \cite{luzzati1967polymorphism, hajduk1994gyroid, batesgyroid, michielsen2008gyroid, almsherqi2012look}. We note that the networks for the gyroid-associated Schwarz primitive ($P$) and diamond ($D$) minimal surfaces \cite{schoen1970infinite} do not exhibit double twist in flat space, although it is possible that the rotation angles between neighboring nodes are so large as to be achiral. Without a reference point between the nodes assignment of twist would be arbitrary.

The $\mathbb{R}^3$ Laves network, by virtue of not filling space continuously, does not require the topological defects typically necessitated by dense double-twisted structures in $\mathbb{R}^3$, such as the blue phase. Nevertheless, the network's double-twist suggests that it may have a natural counterpart in $S^3$ that can elucidate its $\mathbb{R}^3$ properties. Here we construct an analogue to the Laves network in $S^3$, describe its vertices as a regular 24-cell with basis, and detail related spatial constructions for both networks.

\section*{Local Structure of the $S^3$ Laves Network}

\subsection*{$\mathbb{R}^3$ Laves Network Structure}

We have described one construction of the $\mathbb{R}^3$ Laves network but note that
it is also closely related to the face-centered cubic ($\text{fcc}$) lattice. As shown in Figure~\ref{fig:figure_1}a, the $\mathbb{R}^3$ Laves network's vertices lie at $\text{fcc}$ lattice sites; the network edges lie along $\text{fcc}$ nearest-neighbor directions, which correspond to face-diagonal directions of a simple cubic lattice. Because the Wigner-Seitz (Voronoi) cells of the $\text{fcc}$ lattice form the rhombic dodecahedral honeycomb of $\mathbb{R}^3$, each vertex of the $\mathbb{R}^3$ Laves network lies at the center of a rhombic dodecahedral cell of this honeycomb, and its edges perpendicularly intersect three faces of the cell at their face centers. Using the same simple cubic lattice, the Schwarz primitive $P$ network is built from the actual nearest-neighbor edges and the Schwarz diamond $D$ network from the body-diagonals.

\subsection*{$S^3$ Laves Network Structure}

\begin{figure}[ht]
    \centering
     \begin{tabular}{@{}p{0.5\linewidth}@{\quad}p{0.5\linewidth}@{}}
         \stackinset{l}{10pt}{t}{10pt}{(a)}{\includegraphics[width=\linewidth]{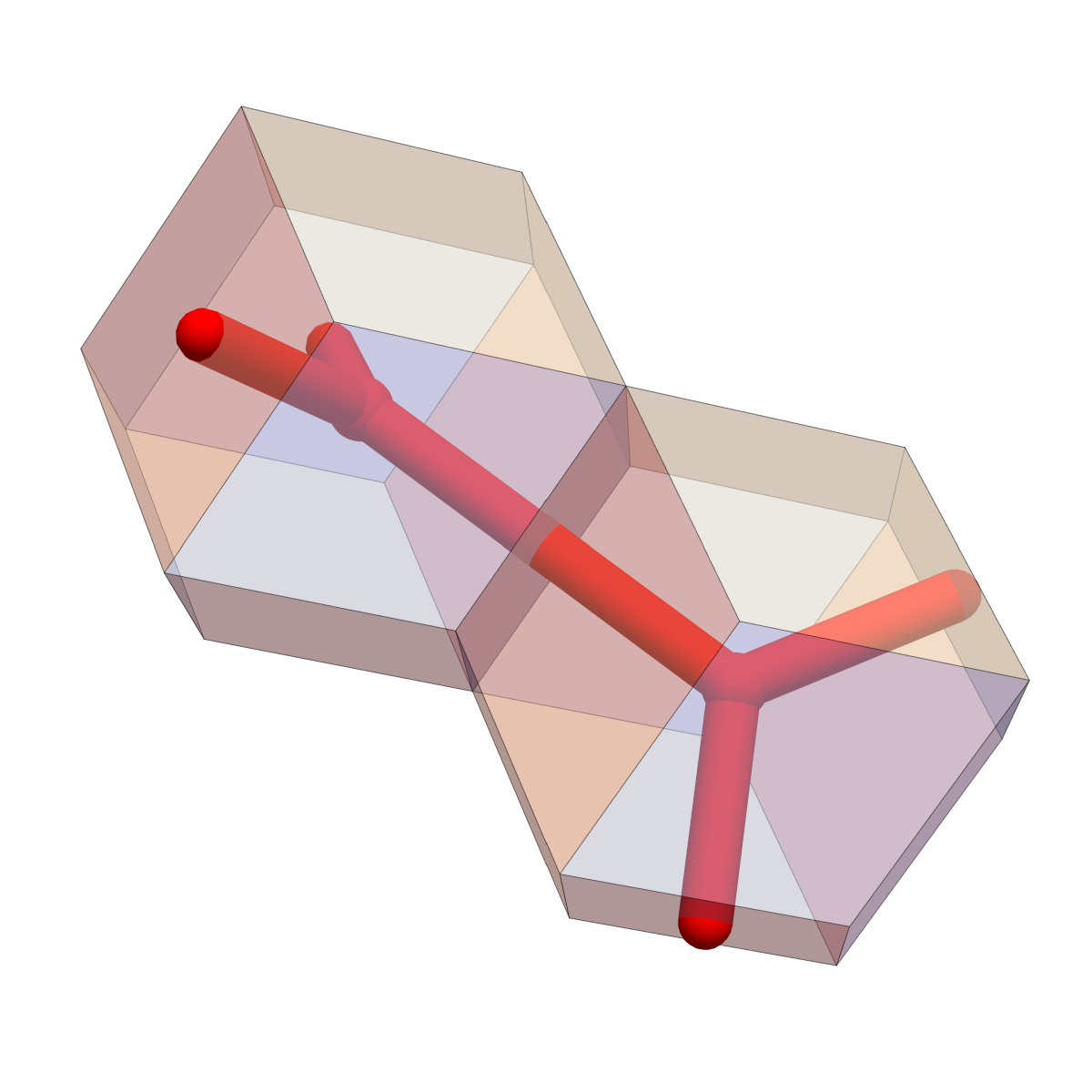}} &
         \stackinset{l}{10pt}{t}{10pt}{(b)}{\includegraphics[width=\linewidth]{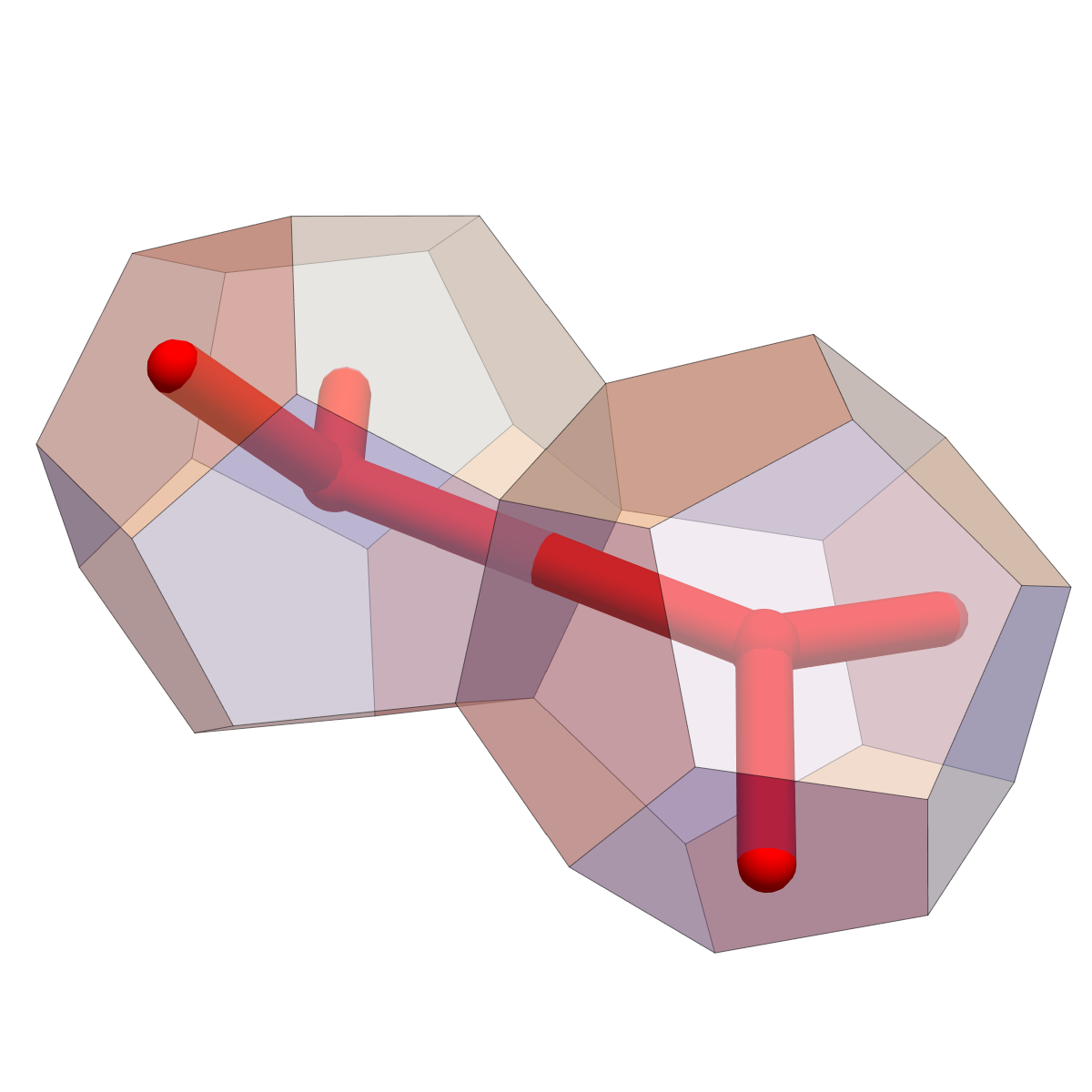}} \\
         \multicolumn{2}{@{}p{\linewidth}@{}}{
            \stackinset{l}{0pt}{t}{0pt}{(c)}{\includegraphics[width=\linewidth]{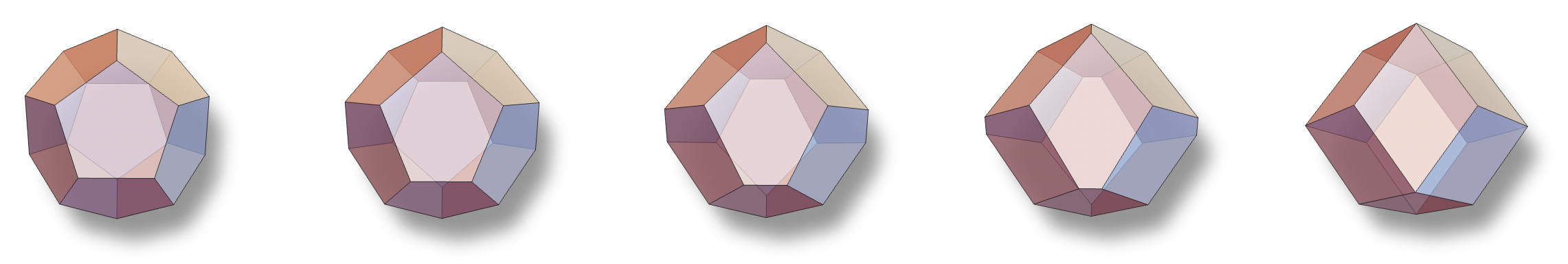}}
         }
    \end{tabular}
     \caption{Local structure of one edge in the $\mathbb{R}^3$ Laves network (a) and the $S^3$ Laves network analogue (b), connected by a transformation that preserves the pyritohedral symmetry of each cell. Rhombic (a) or regular (b) dodecahedral cells may continue to tile the space in $\mathbb{R}^3$ and $S^3$ respectively. (c) The pyritohedral transformation from the rhombic (left) to regular (right) dodecahedron.}
     \label{fig:figure_1}
 \end{figure}

The relationship between the $\mathbb{R}^3$ Laves network structure and the rhombic dodecahedral honeycomb in $\mathbb{R}^3$ suggests a construction method for the desired Laves network analogue in $S^3$, which can be tiled by regular dodecahedra to form the celebrated 120-cell \cite{coxeter1973regular}. A family of pyritohedra connects the rhombic dodecahedron to the regular dodecahedron (Figure~\ref{fig:figure_1}c) \cite{grunbaum1998acoptic}, and correspondingly maps the vertex structure of the $\mathbb{R}^3$ Laves network to a new vertex structure in $S^3$.

From this pyritohedral transformation, we derive the building block of the $S^3$ Laves network as shown in Figure~\ref{fig:figure_1}b. This non-planar ``tripod'' vertex structure is formed by connecting the center of the dodecahedron to the centers of three pentagonal faces that are maximally distant from one another. Although it is non-planar, the $S^3$ tripod vertex structure maintains the three-fold rotational symmetry of its $\mathbb{R}^3$ counterpart. Upon parallel transport, the tripod structure twists by $\pm 4\pi/5$ from vertex to vertex, where the sign of the twist angle remains consistent throughout a single network. Owing to the five-fold symmetry of each face of the dodecahedron, the rotation could have been $\pm 2\pi/5$, but we found that this angle does not result in a network which is both locally Laves-like and non-intersecting. We also note, in passing, that the $S^3$ network can be globally consistently four-colored in exact correspondence with the $\mathbb{R}^3$ network.

Although suggestive, the transformation of local structure from $\mathbb{R}^3$ to $S^3$ cannot be smooth, as the pyritohedral transformations of adjacent cells force opposing rotations of their mutual face. Nevertheless, we can use the resulting edge structure in Figure~\ref{fig:figure_1}b to generate a graph embedded in $S^3$.
Though both rhombic and regular dodecahedral cells tiling $\mathbb{R}^3$ and $S^3$ each have 12 nearest neighbors that are differently-coordinated (forming cuboctaheral and icosahedral graphs respectively), the next-nearest neighbor coordination of an individual cell differs entirely.
Furthermore, neither the rhombic nor regular dodecahedron are inherently chiral objects, but the pyritohedral transformation connecting the two dodecahedra is chiral with respect to the interior vertex structure; however, a consistent chiral transformation of each dodecahedron cannot maintain continuity of the faces between cells, because the faces of each dodecahedron correspond to mirror planes of the dodecahedral honeycomb in both $\mathbb{R}^3$ and $S^3$. Furthermore, only one choice of pyritohedral transformation results in a consistent, non-overlapping network in $S^3$, and this choice corresponds exactly to the chirality of the original network in $\mathbb{R}^3$; \textit{i.e.}, if the network chirality is inverted, then the chirality of the pyritohedral transformation must also be flipped to obtain a consistent, non-intersecting Laves network in $S^3$.

\begin{figure}[ht!]
    \centering
    \begin{tabular}{@{}p{0.4\linewidth}@{\quad}p{0.4\linewidth}@{}}
        \stackinset{l}{10pt}{t}{10pt}{(a)}{\includegraphics[width=\linewidth]{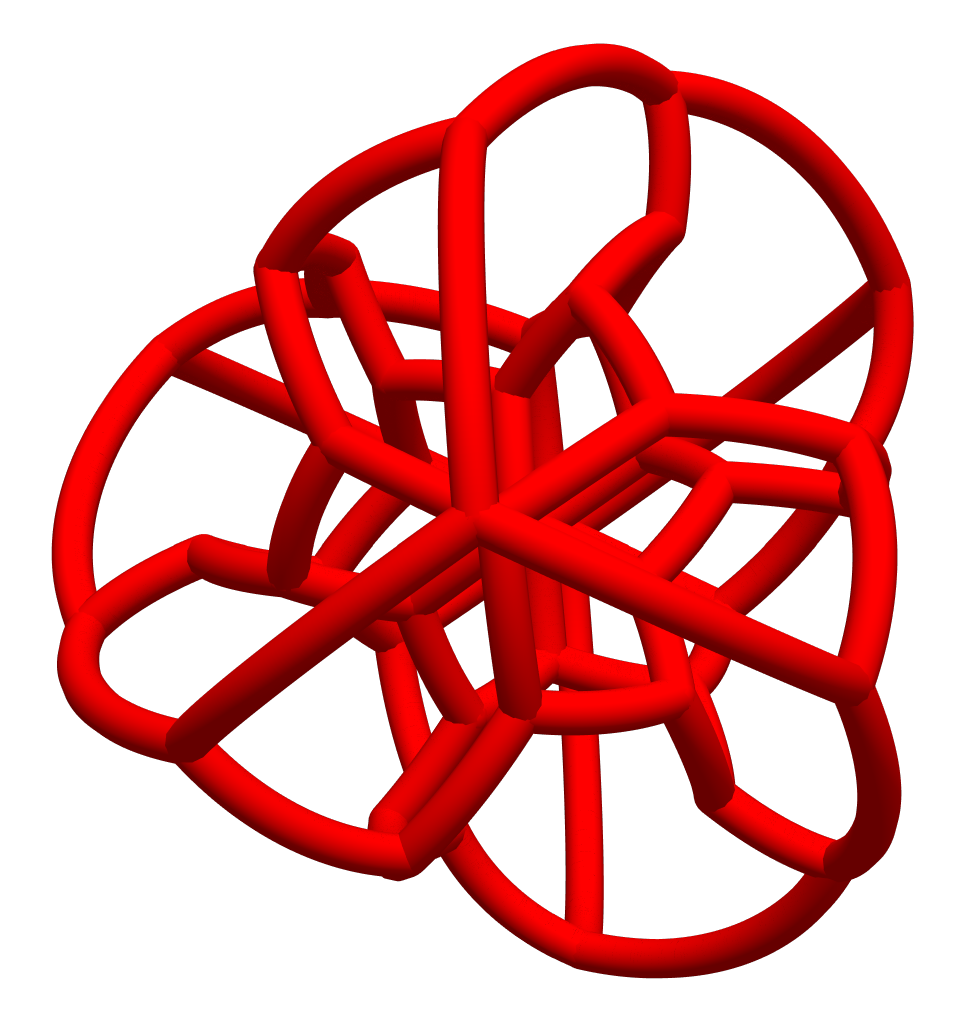}}
        &
        \stackinset{l}{10pt}{t}{10pt}{(b)}{\includegraphics[width=\linewidth]{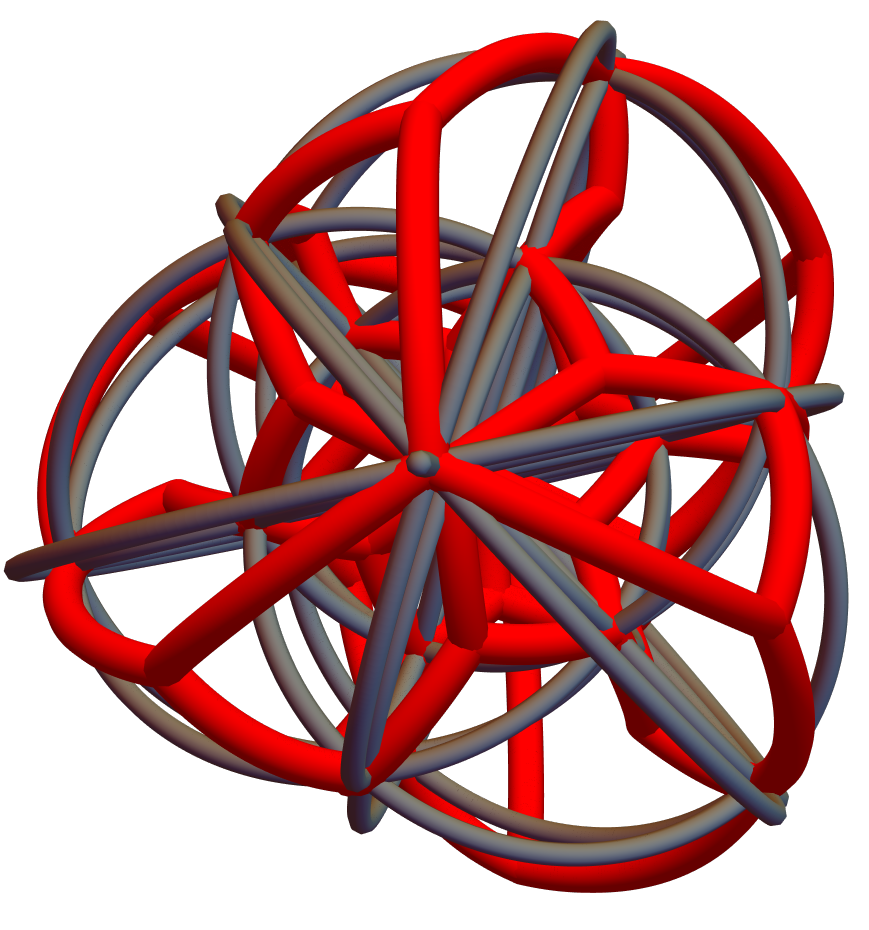}}
    \end{tabular}
    \caption{Projection of the $S^3$ Laves network analogue (red), with an additional view of a 24-cell edges (gray) inscribed in the network. Every other vertex of the $S^3$ Laves network corresponds to a vertex of the 24-cell. The projection function from points on the unit sphere in $\mathbb{R}^4$ is $(x, y, z, w) \mapsto \arccos(w) \left(1-w^2 \right)^{-1/2} (x, y, z)$, with one edge of the 24-cell passing through the point at infinity. All edges follow projections of great circles on the unit sphere in $\mathbb{R}^4$.}
    \label{fig:figure_2}
\end{figure}

\section*{Global Structure of the $S^3$ Laves Network}

From the local structure proposed in Figure~\ref{fig:figure_1}b, the resulting $S^3$ Laves network is vertex- and edge-transitive, consisting of 48 total vertices connected by 72 edges (Figure~\ref{fig:figure_2}a). The network's vertices, each located at the center of a dodecahedron in the 120-cell, form a subset of the 600-cell (a regular 4-dimensional polytope that is dual to the 120 cell) vertices by definition; the network's edges are also subsets of the 600-cell's edges. The shortest loops in the network are of length 8, the ``girth'' of the network. Each network edge is part of 8 such loops, each vertex is part of 12 such loops, and there are 72 such distinct loops in a single network.

As first noted by Coxeter, the flat Laves graph has girth 10 \cite{coxeter1955laves}. That the $S^3$ network has a smaller girth is not surprising: material must somehow be ``redacted'' from $\mathbb{R}^3$ to make the three sphere. Consider a regular dodecahedron in $\mathbb{R}^3$ surrounded by its 12 dodecahedral neighbors. Because the dihedral angle of the dodecahedron is $\arccos\left(-\sqrt{5}/5\right) \approx 116.6^\circ < 120^\circ$, there are gaps between the neighbors. After removing 6 edges and merging their endpoints (Figure~\ref{fig:figure_1}c), we arrive at the rhombic dodecahedron which packs $\mathbb{R}^3$. Thus, we must remove 6 out of 30 edges to correct the dihedral angles of the regular dodecahedron. Each closed loop of length 10 on the $\mathbb{R}^3$ Laves network loses $1/5$ of its bends, resulting in closed loops of length 8 on the $S^3$ Laves network.

Like its flat cousin, the $S^3$ Laves network forms a bipartite graph, where each edge connects vertices from two disjoint subsets of vertices. We note that the 120 vertices of the 600-cell can be partitioned into 5 disjoint sets, each of which is a 24-cell. Given such a partition, the $S^3$ Laves network can be generated by choosing any pair of disjoint 24-cells and connecting their nearest-neighbor vertices.  Accordingly, the $S^3$ Laves network vertices can be viewed as a 24-cell with a two-vertex basis inscribed in a 600-cell (Figure~\ref{fig:figure_2}b).

\begin{figure*}[ht!]
    \centering
    \begin{tabular}{@{}p{0.3\linewidth}@{\quad}p{0.3\linewidth}@{\quad}p{0.3\linewidth}@{}}
        \stackinset{l}{-15pt}{t}{10pt}{(a)}{\includegraphics[width=\linewidth]{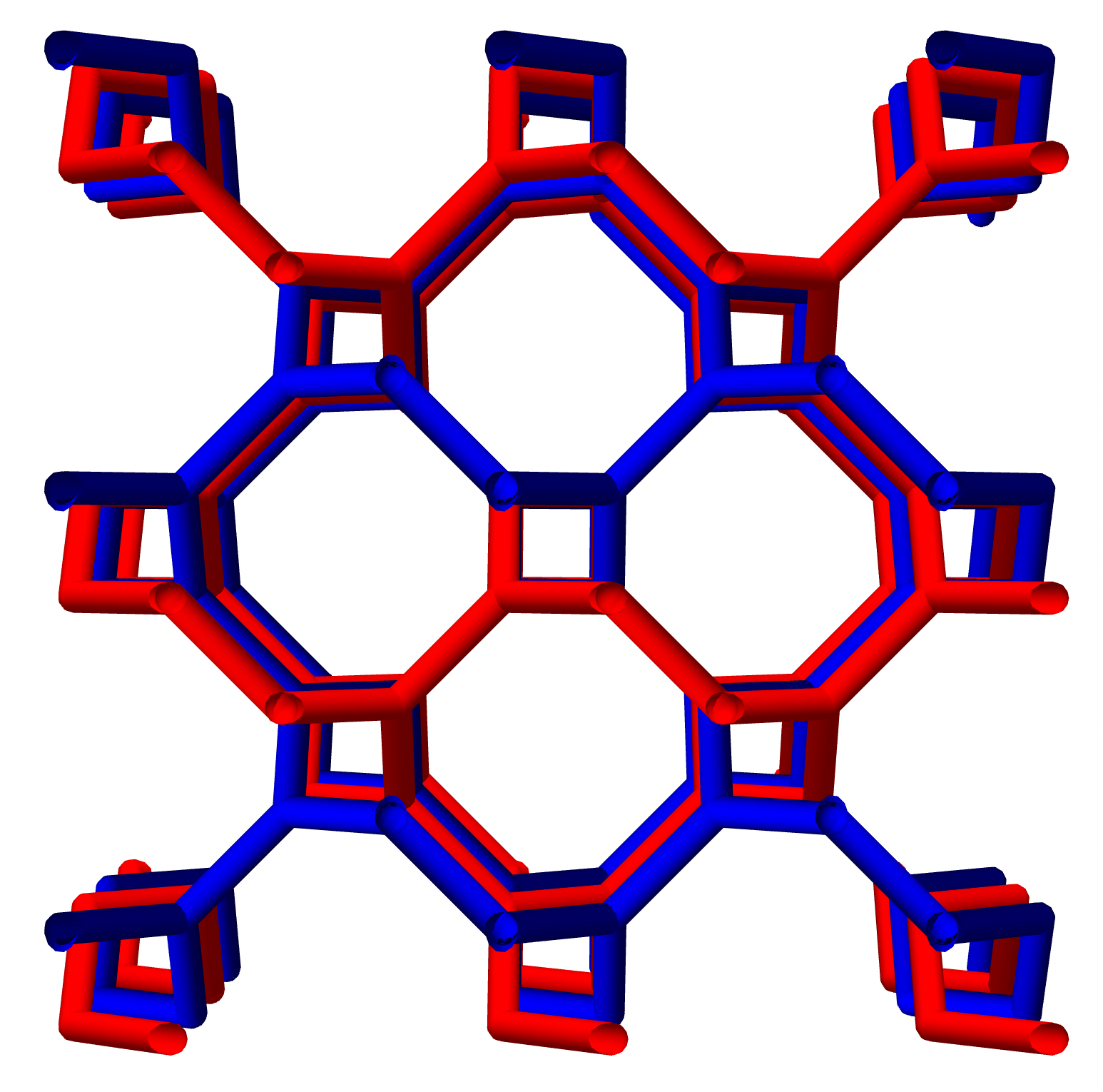}} &
        \stackinset{l}{10pt}{t}{10pt}{(b)}{\includegraphics[width=\linewidth]{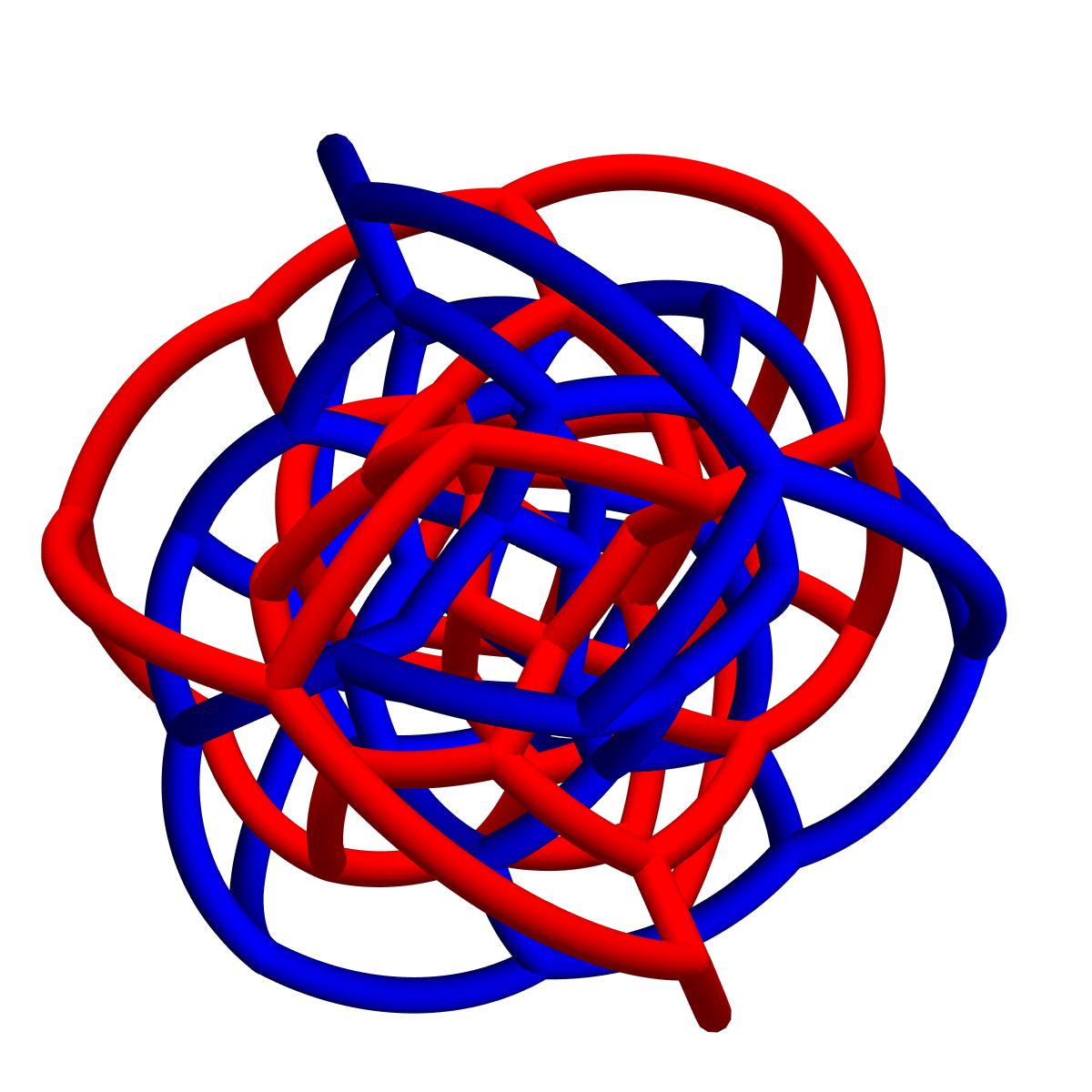}} &
        \stackinset{l}{0pt}{t}{10pt}{(c)}{\includegraphics[width=\linewidth]{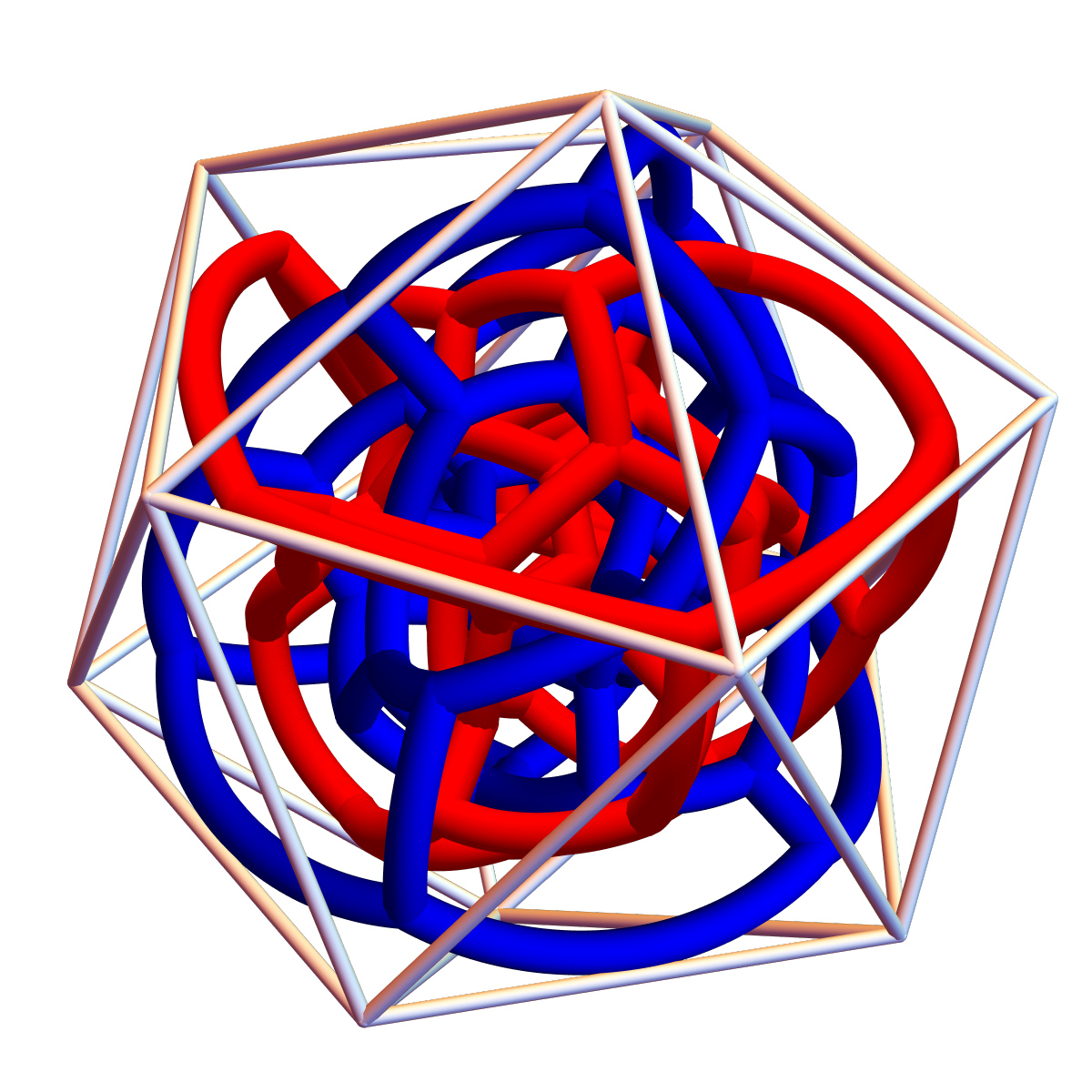}}
    \end{tabular}
    \caption{Two interlaced Laves networks of the same handedness in $\mathbb{R}^3$ (a) and $S^3$ (b); given the twisted edge structure, helical motifs are apparent in both networks. (c) When projected to $\mathbb{R}^3$, the icosahedral structure of the double network becomes apparent; outermost vertices of the double network structure form an icosahedron. Vertices in (b) and (c) are projected from the unit sphere in $\mathbb{R}^4$ to $\mathbb{R}^3$ with the same map as in Figure~\ref{fig:figure_2}, with one vertex of the ``empty'' 24-cell coincident with the point at infinity.}
    \label{fig:figure_3}
\end{figure*}

\section*{Symmetries of the $S^3$ Laves Network}

Although neither the 24-cell nor the 600-cell are themselves chiral objects, the $S^3$ Laves network's vertices (as a union of vertices of two disjoint 24-cells inscribed in the 600-cell) form a chiral structure. Any two disjoint 24-cells are contained in one of ten partitions of the 600-cell's 120 vertices into five disjoint 24-cells; the ten partitions split into two families of five, which are exchanged by reflection symmetries of the 600-cell \cite{denney2020geometry}, and each $S^3$ Laves network inherits chirality from such a partition. Furthermore, the ten choices of partition and $\binom{5}{2} = 10$ choices of constituent 24-cells yields 100 geometrically distinct networks that can be inscribed in a 600-cell, with 50 congruent copies of the network with each chirality.

The symmetry group of the network structure is given by the 144-element maximal subgroup of $H_4$, the 600-cell's symmetry group \cite{koca2006maximal}. Due to the chirality of the $S^3$ Laves network, all of these elements are proper rotations. Geometrically, these symmetries can be understood by decomposing the network's vertices as eight equally spaced discrete Hopf fibers, with each fiber comprising a great hexagon and four fibers making up each constituent 24-cell. Under the Hopf map, these fibers form a cube on the Riemann sphere, allowing 24 rotations interchanging the different fibers. Independently, six rotations can rotate the plane of any great hexagon within the network, yielding 144 isometries of the network structure.

\section*{Interpenetrating Laves Networks in $S^3$}

Since the vertices of the network occupy $48/120 = 2/5$ of the regular dodecahedral lattice sites (contrasted with $1/4$ occupation of rhombic dodecahedral lattice sites in the $\mathbb{R}^3$ Laves network), only a single additional, disjoint network can be constructed based on the same 120-cell honeycomb, and we find that the additional network in $S^3$ must have the same chirality as the original network \footnote{This is very different from the $\mathbb{R}^3$ situation where it is possible to have up to eight catenated networks, four of each handedness \cite{evans2013periodic}.}. Furthermore, because a choice of the first network uniquely encodes a chiral partition of the 600-cell into five disjoint 24-cells, the second network must comprise two other 24-cells from the same partition, leaving a single remaining 24-cell ``empty''. An $S^3$ Laves network with differing chirality could be interlaced with the original network, but only when offset by translations of fractions of a unit cell, so such a mirror network would no longer be made of vertices and edges of the same 600-cell.

Given two interlaced $S^3$ Laves networks constructed from the same 600-cell (as in Figure~\ref{fig:figure_3}b), one might think that the surface dividing them is a spherical version of the gyroid, the minimal surface that divides two interpenetrating $\mathbb{R}^3$ Laves networks of opposite handedness. It should be noted, however, that like \textit{la coupe du Roi} \cite{coupeduroi}, our surface divides $S^3$ into two pieces of the \textit{same} handedness. From the base 120-cell, we construct a discrete version of this surface by removing faces that intersect with the edges of each $S^3$ Laves network, then shrinking each of the dodecahedral cells associated with the remaining ``empty'' 24-cell vertices to a point; leaving 288 faces, 480 edges, and 144 vertices for $\chi = -48$. Equally, we can surround each of the 48 vertices with a 3-punctured sphere ($\chi=-1$) and connect the punctures with cylinders that surround each of the 72 edges ($\chi=0$) producing a manifold with $\chi=-48$. Both constructions imply that the separating surface between two networks of the same chirality has genus 25.  

Although the Lawson minimal surface $\xi_{5,5}$ \cite{lawson1970complete} has the right genus for such a network-separating surface, it has reflection symmetry while the surface we seek must be chiral.  Moreover, given two disjoint $S^3$ Laves networks inscribed on a 600-cell, the 576 rotational symmetries of the ``empty'' 24-cell mix the four remaining disjoint 24-cells that make up the two networks. As only one of every $\binom{4}{2}=6$ such rotations preserves the partition of these four 24-cells into the two specified networks, we conjecture that the dividing surface has a symmetry group of order 96 with no reflections. Whether there exists a genus 25 minimal surface in $S^3$ with this reduced symmetry is unknown to us.

\section*{Discussion}

We have tried to further understand the remarkable Laves graph, the scaffold for the fabled gyroid. Viewing a single network in $\mathbb{R}^3$ as a realization of double twist, we constructed the corresponding network on $S^3$ where this double twist can be accommodated everywhere. The $S^3$ Laves network shares many features of its flat cousin, including a deep relationship with dodecahedral tilings of the ambient space; although, notably, it has girth 8 instead of girth 10 \cite{coxeter1955laves}. A second network can be inscribed in the same underlying 120-cell (or dual 600-cell); but unlike the $\mathbb{R}^3$ Laves network, the second network is the same handedness as the original one.

Does this construction give us insight into nature's preference for the gyroid over the Schwarz primitive ($P$) and diamond ($D$) minimal surface morphologies? Though they do not obviously have double twist, the corresponding $P$ and $D$ networks can also be embedded in $S^3$ via the edge structures of the 8-cell and the 16-cell, respectively. The Laves graph requires a larger structure than either of these and thus requires less distortion to be stretched into flat space. However, we are not aware of any argument that would suggest that the $S^3$ network is optimal in any of the ways the gyroid network is \cite{schroder2006bicontinuous}.

\section*{Materials and Methods}

All computations were performed either in Mathematica (Wolfram), with Zometool kits, or by pure thought.
% Code used to generate figures is available as Supplementary Information.
The use of AI tools was assiduously avoided.

\section*{Acknowledgments}

The authors thank Myfwany Evans, Stephen Hyde, and Matthias Weber for insightful discussions and Reiko Schoen for supplying us with original source models of the gyroid. Mark Bowick introduced us to \textit{la coupe du Roi}. This work was supported by a Simons Investigatorship to R.D.K.

\bigskip

\bibliographystyle{unsrt}
\bibliography{references}

@book{euclid-300elements,
  author = {Euclid},
  title = {The Elements},
  year = {300 BC}
}

@book{bragg1915x,
  title={X-rays and Crystal Structure},
  author={Bragg, William Henry and Bragg, William Lawrence},
  volume={215},
  year={1915},
  publisher={G. Bell and Sons, Ltd.}
}

@article{laves1932klassifikation,
  title={Zur {K}lassifikation der {S}ilikate. {G}eometrische {U}ntersuchungen M{\"o}glicher {S}ilicium-{S}auerstoff-{V}erb{\"a}nde als {V}erkn{\"u}pfungsm{\"o}glichkeiten Regul{\"a}rer {T}etraeder},
  author={Laves, Fritz},
  journal={Zeitschrift f{\"u}r Kristallographie-Crystalline Materials},
  volume={82},
  number={1-6},
  pages={1--14},
  year={1932},
  publisher={De Gruyter Oldenbourg}
}

@book{brillouin1953wave,
  title={Wave propagation in periodic structures: electric filters and crystal lattices},
  author={Brillouin, L{\'e}on Nicholas},
  publisher={McGraw-Hill Book Company, Inc.},
  year={1953}
}

@article{coxeter1955laves,
  title={On {L}aves' graph of girth ten},
  author={Coxeter, H. S. M.},
  journal={Canadian Journal of Mathematics},
  volume={7},
  pages={18--23},
  year={1955},
  publisher={Cambridge University Press}
}

@article{coxeter1958close,
  title={Close-packing and froth},
  author={Coxeter, H. S. M.},
  journal={Illinois Journal of Mathematics},
  volume={2},
  number={4B},
  pages={746--758},
  year={1958},
  publisher={Duke University Press}
}

@article{frank1958complex,
  title={Complex Alloy Structures Regarded as Sphere Packings. {I}. Definitions and Basic Principles},
  author={Frank, Frederick C. and Kasper, John S.},
  journal={Acta Crystallographica},
  volume={11},
  number={3},
  pages={184--190},
  year={1958},
  publisher={International Union of Crystallography}
}

@article{frank1959complex,
  title={Complex Alloy Structures Regarded as Sphere Packings. {II}. Analysis and Classification of Representative Structures},
  author={Frank, Frederick C. and Kasper, John S.},
  journal={Acta Crystallographica},
  volume={12},
  number={7},
  pages={483--499},
  year={1959},
  publisher={International Union of Crystallography}
}

@article{luzzati1967polymorphism,
  title={Polymorphism of lipids},
  author={Luzzati, Vittorio and Spegt, P. A.},
  journal={Nature},
  volume={215},
  number={5102},
  pages={701--704},
  year={1967},
  publisher={Nature Publishing Group UK London}
}

@article{lawson1970complete,
  title={Complete minimal surfaces in {$S^3$}},
  author={Lawson, H. Blaine},
  journal={Annals of Mathematics},
  volume={92},
  number={3},
  pages={335--374},
  year={1970},
  publisher={JSTOR}
}

@book{schoen1970infinite,
  title={Infinite periodic minimal surfaces without self-intersections},
  author={Schoen, Alan Hugh},
  volume={5541},
  year={1970},
  publisher={National Aeronautics and Space Administration}
}

@article{wilson1972critical,
  title={Critical exponents in 3.99 dimensions},
  author={Wilson, Kenneth G. and Fisher, Michael E.},
  journal={Physical Review Letters},
  volume={28},
  number={4},
  pages={240},
  year={1972},
  publisher={APS}
}

@article{coates1973optical,
  title={Optical studies of the amorphous liquid-cholesteric liquid crystal transition: The “blue phase”},
  author={Coates, D. and Gray, G. W.},
  journal={Physics Letters A},
  volume={45},
  number={2},
  pages={115--116},
  year={1973},
  publisher={Elsevier}
}

@book{coxeter1973regular,
  title={Regular polytopes},
  author={Coxeter, H. S. M.},
  year={1973},
  publisher={Courier Corporation}
}

@article{sethna1983relieving,
  title={Relieving cholesteric frustration: the blue phase in a curved space},
  author={Sethna, James P. and Wright, David C. and Mermin, N. D.},
  journal={Physical review letters},
  volume={51},
  number={6},
  pages={467},
  year={1983},
  publisher={APS}
}

@article{nelson1989polytetrahedral,
  title={Polytetrahedral order in condensed matter},
  author={Nelson, David R. and Spaepen, Frans},
  journal={Solid State Physics},
  volume={42},
  pages={1--90},
  year={1989},
  publisher={Elsevier}
}

@article{wright1989crystalline,
  title={Crystalline liquids: the blue phases},
  author={Wright, David C. and Mermin, N. D.},
  journal={Reviews of Modern physics},
  volume={61},
  number={2},
  pages={385},
  year={1989},
  publisher={APS}
}

@article{hajduk1994gyroid,
  title={The gyroid: a new equilibrium morphology in weakly segregated diblock copolymers},
  author={Hajduk, Damian A. and Harper, Paul E. and Gruner, Sol M. and Honeker, Christian C. and Kim, Gia and Thomas, Edwin L. and Fetters, Lewis J.},
  journal={Macromolecules},
  volume={27},
  number={15},
  pages={4063--4075},
  year={1994},
  publisher={ACS Publications}
}

@article{batesgyroid,
  title={Epitaxial relationship for hexagonal-to-cubic phase transition in a book copolymer mixture},
  author={Schulz, Mark F. and Bates, Frank S. and Almdal, Kristoffer and Mortensen, Kell},
  journal={Physical review letters},
  volume={73},
  number={1},
  pages={86},
  year={1994},
  publisher={APS}
}

@article{grunbaum1998acoptic,
  title={Acoptic polyhedra},
  author={Gr{\"u}nbaum, Branko},
  journal={Contemporary Mathematics},
  volume={223},
  pages={163--200},
  year={1998},
  publisher={American Mathematical Society}
}

@book{sadoc1999geometrical,
  title={Geometrical frustration},
  author={Sadoc, Jean-Fran{\c{c}}ois and Mosseri, Rimy},
  publisher={Cambridge University Press},
  year={1999}
}

@article{koca2006maximal,
  title={Maximal subgroups of the {C}oxeter group {$W(H_4$)} and quaternions},
  author={Koca, Mehmet and Ko{\c{c}}, Ramazan and Al-Barwani, Muataz and Al-Farsi, Shadia},
  journal={Linear algebra and its applications},
  volume={412},
  number={2-3},
  pages={441--452},
  year={2006},
  publisher={Elsevier}
}

@article{schroder2006bicontinuous,
  title={Bicontinuous geometries and molecular self-assembly: comparison of local curvature and global packing variations in genus-three cubic, tetragonal and rhombohedral surfaces},
  author={Schr{\"o}der-Turk, Gerd E. and Fogden, Andrew and Hyde, Stephen T.},
  journal={The European Physical Journal B},
  volume={54},
  number={4},
  pages={509--524},
  year={2006},
  publisher={Springer}
}

@article{virial,
  title={Hard disks on the hyperbolic plane},
  author={Modes, Carl D. and Kamien, Randall D.},
  journal={Physical review letters},
  volume={99},
  number={23},
  pages={235701},
  year={2007},
  publisher={APS}
}

@article{hyde2008short,
  title={A short history of an elusive yet ubiquitous structure in chemistry, materials, and mathematics},
  author={Hyde, Stephen T. and O'Keeffe, Michael and Proserpio, Davide M.},
  journal={Angewandte Chemie International Edition},
  volume={47},
  number={42},
  pages={7996--8000},
  year={2008},
  publisher={Wiley-VCH Verlag}
}

@article{michielsen2008gyroid,
  title={Gyroid cuticular structures in butterfly wing scales: biological photonic crystals},
  author={Michielsen, Kristel and Stavenga, Doekele G.},
  journal={Journal of The Royal Society Interface},
  volume={5},
  number={18},
  pages={85--94},
  year={2008},
  publisher={The Royal Society}
}

@article{modes2008geometrical,
  title={Geometrical frustration in two dimensions: Idealizations and realizations of a hard-disk fluid in negative curvature},
  author={Modes, Carl D. and Kamien, Randall D.},
  journal={Physical Review E},
  volume={77},
  number={4},
  pages={041125},
  year={2008},
  publisher={APS}
}

@article{almsherqi2012look,
  title={A look through ‘lens’ cubic mitochondria},
  author={Almsherqi, Zakaria and Margadant, Felix and Deng, Yuru},
  journal={Interface focus},
  volume={2},
  number={5},
  pages={539--545},
  year={2012},
  publisher={The Royal Society}
}

@article{evans2013periodic,
author = "Evans, Myfanwy E. and Robins, Vanessa and Hyde, Stephen T.",
title = "{Periodic entanglement {I}: networks from hyperbolic reticulations}",
journal = "Acta Crystallographica Section A",
year = "2013",
volume = "69",
number = "3",
pages = "241--261",
month = "May",
doi = {10.1107/S0108767313001670},
}

@article{coupeduroi,
  title={\textit{La coupe du roi} and other methods to halve objects},
  author={Schwarzenbach, Dieter},
  journal={Zeitschrift f{\"u}r Kristallographie-Crystalline Materials},
  volume={230},
  number={12},
  pages={761--765},
  year={2015},
  publisher={De Gruyter}
}

@article{denney2020geometry,
  title={The geometry of {$H_4$} polytopes},
  author={Denney, Tomme and Hooker, Da’Shay and Johnson, De’Janeke and Robinson, Tianna and Butler, Majid and Claiborne, Sandernisha},
  journal={Advances in Geometry},
  volume={20},
  number={3},
  pages={433--444},
  year={2020},
  publisher={De Gruyter}
}

@book{o2020crystal,
  title={Crystal structures},
  author={O'Keeffe, Michael and Hyde, Bruce G.},
  year={2020},
  publisher={Courier Dover Publications}
}

@article{schonhofer2023rationalizing,
  title={Rationalizing {E}uclidean assemblies of hard polyhedra from tessellations in curved space},
  author={Sch{\"o}nh{\"o}fer, Philipp W. A. and Sun, Kai and Mao, Xiaoming and Glotzer, Sharon C.},
  journal={Physical Review Letters},
  volume={131},
  number={25},
  pages={258201},
  year={2023},
  publisher={APS}
}

\end{document}